\documentclass[12pt,a4paper]{article} 
\usepackage[utf8]{inputenc}
\usepackage{latexsym}
\usepackage{amssymb}
\usepackage{amsmath}
\usepackage{amsthm}
\usepackage{graphicx}
\usepackage{bm}
\usepackage{verbatim}
\usepackage[breaklinks,backref]{hyperref}

\newtheorem{lemma}{Lemma}
\newtheorem{theorem}{Theorem}

\newtheorem{ourclaim}{Claim}
\theoremstyle{definition}
\newtheorem{definition}{Definition}

\newtheorem{oldtheorem}{Theorem}

\newcommand{\set}[2]{\ensuremath{ \{ \, #1 \mid #2 \, \} }}

\renewcommand{\emptyset}{\varnothing}
\renewcommand{\epsilon}{\varepsilon}

\newcommand{\rank}{\mathop{\mathrm{rank}}}
\newcommand{\Sstart}{S_{\mathrm{start}}}
\newcommand{\Hstart}{H_{\mathrm{start}}}
\newcommand{\Hfake}{H_{\mathrm{dead\,end}}}
\newcommand{\vexit}{v_{\mathrm{exit}}}

\begin{document}

\sloppy

\title{Homomorphisms on graph-walking automata\thanks{%
	This work was supported by
	the Ministry of Science and Higher Education of the Russian Federation,
	agreement 075-15-2019-1619.
	}}
\author{Olga Martynova\thanks{%
	Department of Mathematics and Computer Science,
	St.~Petersburg State University, 7/9 Universitetskaya nab., Saint Petersburg 199034, Russia,
	\emph{and}
	Leonhard Euler International Mathematical Institute at St. Petersburg State University,
	Saint Petersburg, Russia.
	E-mail:
	\texttt{olga22mart@gmail.com}.}
	\and
	Alexander Okhotin\thanks{%
	Department of Mathematics and Computer Science,
	St.~Petersburg State University, 7/9 Universitetskaya nab., Saint Petersburg 199034, Russia.
	E-mail:
	\texttt{alexander.okhotin@spbu.ru}.}
}

\maketitle

\begin{abstract}
Graph-walking automata (GWA) are a model for graph traversal using finite-state control:
these automata move between the nodes of an input graph, following its edges.
This paper investigates the effect of node-replacement graph homomorphisms
on recognizability by these automata.
It is not difficult to see that
the family of graph languages recognized by GWA
is closed under inverse homomorphisms.
The main result of this paper is that, for $n$-state automata
operating on graphs with $k$ labels of edge end-points,
the inverse homomorphic images require GWA
with $kn+O(1)$ states in the worst case.
The second result is that already for tree-walking automata,
the family they recognize is not closed under injective homomorphisms.
Here the proof is based on an easy homomorphic characterization of regular tree languages.
\end{abstract}

\sloppy

\section{Introduction}

A graph-walking automaton moves over a labelled graph using a finite set of states
and leaving no marks on the graph.
This is a model of a robot finding its way in a maze.
There is a classical result by Budach~\cite{Budach}
that for every automaton there is a graph in which it cannot visit all nodes,
see a modern proof by Fraigniaud et al.~\cite{FraigniaudIlcinkasPeerPelcPeleg}.
On the other hand, Disser et al.~\cite{DisserHackfeldKlimm}
recently proved that if such an automaton
is additionally equipped with $O(\log \log n)$ memory
and $O(\log \log n)$ pebbles,
then it can traverse every graph with $n$ nodes,
and this amount of resources is optimal.
For graph-walking automata,
there are results on the construction of halting and reversible automata
by Kunc and Okhotin~\cite{KuncOkhotin_reversible},
as well as recent lower bounds on the complexity of these transformations
established by the authors~\cite{MartynovaOkhotin_lb}.

Graph-walking automata are a generalization of two-way finite automata
and tree-walking automata.
Two-way finite automata are a standard model in automata theory,
and the complexity of their determinization remains a major open problem,
notable for its connection to the L vs. NL problem in the complexity theory~\cite{Kapoutsis_logspace}.
Tree-walking automata (TWA) have received particular attention in the last two decades,
with important results on their expressive power
established by Boja\'nczyk and Colcombet~\cite{BojanczykColcombet_det,BojanczykColcombet_reg}.

The theory of tree-walking and graph-walking automata needs further development.
In particular, not much is known about their size complexity.
For two-way finite automata (2DFA),
only the complexity of transforming them to one-way automata
has been well researched~\cite{Kapoutsis,GeffertMereghettiPighizzini2003,TwoWayDFAs}.
Also there are some results
on the complexity of operations on 2DFA~\cite{JiraskovaOkhotin_2dfa,two_way_dfa_sc_2},
which also rely on the transformation to one-way automata.
These proof methods have no analogues for TWA and GWA,
and the complexity of operations on these models remains uninvestigated.
The lower bounds on the complexity of transforming
graph-walking automata to halting and reversible~\cite{MartynovaOkhotin_lb}
in turn have no analogues for TWA and 2DFA.

This paper continues the investigation
of the state complexity of graph-walking automata,
with some results extending to tree-walking automata.
The goal is to study some of the few available operations on graphs:
node-replacement homomorphisms, as well as inverse homomorphisms.
In the case of strings, a homomorphism is defined by the identity $h(uv)=h(u)h(v)$,
and the class of regular languages is closed under all homomorphisms,
as well as under their inverses, defined by $h^{-1}(L)=\set{w}{h(w) \in L}$.
For the 2DFA model, the complexity of inverse homomorphisms is known:
as shown by Jir\'askov\'a and Okhotin~\cite{JiraskovaOkhotin_2dfa},
it is exactly $2n$ in the worst case, where $n$ is the number of states in the original automaton.
However, this proof is based on the transformations between one-way and two-way finite automata,
which is a property unique for the string case.
The state complexity of homomorphisms for 2DFA is known to lie
between exponential and double exponential~\cite{JiraskovaOkhotin_2dfa}.
For tree-walking and graph-walking automata,
no such questions were investigated before,
and they are addressed in this paper.

The closure of graph-walking automata under every inverse homomorphism
is easy:
in Section~\ref{section_invh} it is shown that,
for an $n$-state GWA, there is a GWA with $nk+1$ states for its inverse homomorphic image,
where $k$ is the number of labels of edge end-points.
If the label of the initial node is unique, then $nk$ states are enough.
This transformation is proved to be optimal
by establishing a lower bound of $nk$ states.
The proof of the lower bound makes use of a graph
that is easy to pass in one direction and hard to pass in reverse,
constructed in the authors'~\cite{MartynovaOkhotin_lb} recent paper.

The other result of this paper,
presented in Section~\ref{section_reg_tree},
is that the family of tree languages recognized by tree-walking automata
is not closed under injective homomorpisms,
thus settling this question for graph-walking automata as well.
The result is proved by first establishing
a characterization of regular tree languages
by a combination of an injective homomorphism and an inverse homomorphism.
This characterization 
generalizes a known result by Latteux and Leguy~\cite{LatteuxLeguy},
see also an earlier result by \v{C}ul\'{\i}k et al.~\cite{CulikFichSalomaa}.
In light of this characterization, a closure under injective homomorphisms
would imply that every regular tree language is recognized by a tree-walking automaton,
which would contradict the famous result
by Boja\'nczyk and Colcombet~\cite{BojanczykColcombet_reg}.

\section{Graph-walking automata}\label{section_definitions}

Formalizing the definition of graph-walking automata (GWA)
requires a more elaborate notation than
for 2DFA and TWA.
It begins with a generalization of an alphabet to the case of graphs:
a \emph{signature}.

\begin{definition}[Kunc and Okhotin~\cite{KuncOkhotin_reversible}]
A signature $S$ is a quintuple $S = (D, -, \Sigma, \Sigma_0, (D_a)_{a \in \Sigma})$,
where:
\begin{itemize}
\item
	$D$ is a finite set of directions,
	which are labels attached to edge end-points;
\item
	a bijection $- \colon D \to D$
	provides an opposite direction,
	with $-(-d) = d$ for all $d \in D$;
\item
	$\Sigma$ is a finite set of node labels;
\item
	$\Sigma_0 \subseteq \Sigma$
	is a non-empty subset of possible labels of the initial node;
\item
	$D_a \subseteq D$, for every label $a \in \Sigma$,
	is the set of directions in nodes labelled with $a$.
\end{itemize}
\end{definition}

Like strings are defined over an alphabet,
graphs are defined over a signature.

\begin{definition}
A graph over a signature $S = (D, -, \Sigma, \Sigma_0, (D_a)_{a \in \Sigma})$
is a quadruple $(V, v_0, +, \lambda)$, where:
\begin{itemize}
\item
	$V$ is a finite set of nodes;
\item
	$v_0 \in V$ is the initial node;
\item
	edges are defined by a partial function $+ \colon V \times D \to V$, such that
	if $v+d$ is defined, then $(v+d) + (-d)$ is defined and equals $v$;
\item a total mapping $\lambda \colon V \to \Sigma$, such that
	$v+d$ is defined
	if and only if
	$d \in {D_{\lambda(v)}}$,
	and
	$\lambda(v) \in \Sigma_0$
	if and only if
	$v = v_0$.
\end{itemize}
The set of all graphs over $S$ is denoted by $L(S)$.
\end{definition}

In this paper, all graphs are finite and connected.

A graph-walking automaton is defined similarly to a 2DFA,
with an input graph instead of an input string.

\begin{definition}
\emph{A (deterministic) graph-walking automaton (GWA) over a signature 
$S = (D, -, \Sigma, \Sigma_0, (D_a)_{a \in \Sigma})$} 
is a quadruple $A = (Q, q_0, F, \delta)$, where
\begin{itemize}
\item $Q$ is a finite set of states;
\item $q_0 \in Q$ is the initial state;
\item $F \subseteq Q \times \Sigma$ is a set of acceptance conditions;
\item $\delta \colon (Q \times \Sigma) \setminus F \to Q \times D$ is 
a partial transition function, with $\delta(q, a) \in Q \times D_a$ for all $a$ 
and $q$ where $\delta$ is defined.
\end{itemize}
A computation of a GWA on a graph $(V, v_0, +, \lambda)$
is a uniquely defined
sequence of configurations $(q, v)$, with $q \in Q$ and $v \in V$.
It begins with $(q_0, v_0)$
and proceeds from $(q, v)$ to $(q', v+d)$, where $\delta(q, \lambda(v))=(q', d)$.
The automaton accepts by reaching $(q, v)$ with $(q, \lambda(v)) \in F$.

On each input graph, a GWA can accept, reject or loop.
The set of all graphs accepted is denoted by $L(A)$.
\end{definition}


The operation on graphs investigated in this paper
is a homomorphism that replaces nodes with subgraphs.

\begin{definition}
[Graph homomorphism]
Let $S$ and $\widehat{S}$ be two signatures,
with the set of directions of $S$ contained in the set of directions of $\widehat{S}$.
A mapping $h \colon L(S) \to L(\widehat{S})$ is a (node-replacement) homomorphism,
if, for every graph $G$ over $S$,
the graph $h(G)$ is constructed out of $G$ as follows.
For every node label $a$ in $S$,
there is a connected subgraph $h(a)$ over the signature $\widehat{S}$,
which has an edge leading outside for every direction in $D_a$;
these edges are called \emph{external}.
Then, $h(G)$ is obtained out of $G$ by replacing every node $v$ with a subgraph $h(v)=h(a)$,
where $a$ is the label of $v$, so that the edges that come out of $v$ in $G$
become the external edges of this copy of $h(a)$.

The subgraph $h(a)$ must contain at least one node.
It contains an initial node if and only if the label $a$ is initial.
\end{definition}

\section{Inverse homomorphisms: upper and lower bounds}\label{section_invh}

Given a graph-walking automaton $A$ and a homomorphism $h$,
the inverse homomorphic image $h^{-1}(L(A))$
can be recognized by another automaton
that, on a graph $G$,
simulates the operation of $A$ on the image $h(G)$.
A construction of such an automaton is presented in the following theorem.

\begin{theorem}
\label{theorem_inverse_homomorphism_upper_bound}
Let $S$ be a signature with $k \geqslant 1$ directions,
and let $\widehat{S}$ be a signature containing all directions from $S$.
Let $h \colon L(S) \to L(\widehat{S})$ be a graph homomorphism between these signatures.
Let $A$ be a graph-walking automaton with $n$ states
that operates on graphs over $\widehat{S}$.
Then there exists a graph-walking automaton $B$ with $nk+1$ states,
operating on graphs over $S$, 
which accepts a graph $G$
if and only if $A$ accepts its image $h(G)$.
If $S$ contains a unique initial label,
then it is sufficient to use $nk$ states.
\end{theorem}

In order to carry out the simulation of $A$ on $h(G)$
while working on $G$,
it is sufficient for $B$ to remember the current state of $A$
and the direction in which $A$ has entered the image in $h(G)$ of the current node of $B$.

\begin{proof}
Let the first signature be $S = (D, -, \Sigma, \Sigma_0, (D_a)_{a \in \Sigma})$.
Let $A = (Q, q_0, F, \delta)$.
The new automaton is defined as $B = (P, p_0, E, \sigma)$.

When $B$ operates on a graph $G$,
it simulates the computation of $A$ on $h(G)$.
The set of states of $B$ is $P=(Q \times D) \cup \{p_0\}$,
where $p_0$ is a non-reenterable initial state;
if there is only one initial label in $S$, then the state $p_0$ is omitted.
All other states in $B$ are of the form $(q, d)$,
where $q$ is a state of $A$,
and $d$ is a direction in $G$.
When $B$ is at a node $v$ in a state $(q,d)$,
it simulates $A$ having entered the subgraph $h(v)$ from the direction $d$ in the state $q$.

The transition function $\sigma$ and the set of accepting states $E$ of $B$
are defined by simulating $A$ on subgraphs.
For a state of the form $(q,d)$, and for every label $a \in \Sigma$, with $-d \in D_a$, 
the goal is to decide whether $((q,d),a)$ is an accepting pair,
and if not, then what is the transition $\sigma((q,d),a)$.
To this end, the automaton $A$ is executed on the subgraph $h(a)$,
entering this subgraph in the direction $d$ in the state $q$.
If $A$ accepts without leaving $h(a)$, then the pair $((q,d),a)$ is defined as accepting in $B$.
Otherwise, if $A$ rejects or loops inside $h(a)$, 
then $\sigma((q,d),a)$ is left undefined.
If $A$ leaves $h(a)$ by an external edge in the direction $d'$ in a state $q'$,
then $B$ has a transition $\sigma((q,d),a) = ((q',d'), d')$.

If $S$ has a unique initial label, $\Sigma_0=\{a_0\}$,
then the automaton $A$ always starts in the subgraph $h(a_0)$,
and its initial state can be defined by the same method as above,
by considering the computation of $A$ on this subgraph
starting in the initial state at the initial node.
If $A$ accepts, rejects or loops without leaving the subgraph $h(a_0)$,
then it is sufficient to have $B$ with a single state, in which it gives an immediate answer.
If $A$ leaves the subgraph in the direction $d$, changing from $q$ to a state $q'$,
then the state $(q, -d)$ can be taken as the initial state of $B$;
then $B$ starts simulating the computation of $A$ from this point.

If there are multiple initial labels in $\Sigma_0$,
then the automaton $B$ uses a separate initial state $p_0$.
The transitions in $p_0$ and its accepting status
are defined only on initial labels, as follows.
Let $a_0 \in \Sigma_0$ be an initial label,
and consider the computation of $A$ on the subgraph $h(a_0)$,
starting at the initial node therein, in the initial state.
If $A$ accepts inside $h(a_0)$,
then $(p_0,a_0)$ is an accepting pair.
Otherwise, if $A$ rejects or loops without leaving $h(a_0)$,
then $\sigma(p_0,a_0)$ is not defined.
If $A$ leaves $h(a_0)$ in the direction $d'$ in the state $q'$,
then the transition is $\sigma(p_0,a_0) = ((q', d'),d')$.

The automaton $B$ has $nk$ or $nk+1$ states,
and it operates over $S$.
The following correctness claim for this construction
can be proved by induction on the number of steps made by $B$ on $G$.

\begin{ourclaim} \label{upper_bound_induction_step}
Assume that the automaton $B$,
after $t \geqslant 1$ steps of its computation on $G$,
is in a state $(q',d')$ at a node $v$.
Then, in the computation of $A$ on $h(G)$
there is a moment $\widehat{t} \geqslant t$, 
at which $A$ enters the subgraph $h(v)$ 
in the direction $d'$ in the state $q'$
(the only exception is the initial state of $B$ in the case $p_0$ is not used).
\end{ourclaim}

It follows that the automaton $B$ thus defined
indeed accepts a graph $G$ if and only if $A$ accepts $h(G)$.
\end{proof}

It turns out that this expected construction is actually optimal,
as long as the initial label is unique:
the matching lower bound of $nk$ states is proved below.

\begin{theorem}
\label{theorem_inverse_homomorphism_lower_bound} 
For every $k \geqslant 9$,
there is a signature $S$ with $k$ directions
and a homomorphism $h \colon L(S) \to L(S)$,
such that for every $n \geqslant 4$,
there exists an $n$-state automaton $A$ over the signature $S$, 
such that every automaton $B$,
which accepts $G$ if and only if $A$ accepts $h(G)$,
has at least $nk$ states.
\end{theorem}

Proving lower bounds on the size of graph-walking automata is generally not easy.
Informally, it has to be proved that the automaton must remember a lot;
however, in theory, it can always return to the initial node
and recover all the information is has forgotten.
In order to eliminate this possibility,
the initial node shall be placed in a special subgraph $\Hstart$,
from which the automaton can easily get out,
but if it ever needs to reenter this subgraph,
finding the initial node would require too many states.
This subgraph is constructed in the following lemma;
besides $\Hstart$, there is another subgraph $\Hfake$,
which is identical to $\Hstart$ except for not having an initial label;
then, it would be hard for an automaton to distinguish between these two subgraphs from the outside,
and it would not identify the one in which it has started.

\begin{lemma} \label{lemma_H_start_H_fake}
For every $k \geqslant 4$ there is a signature $\Sstart$ with $k$ directions,
with two pairs of opposite directions $a$, $-a$ and $b$, $-b$,
such that for every $n \geqslant 2$
there are two graphs $\Hstart$ and $\Hfake$ over this signature,
with the following properties.
\begin{enumerate}\renewcommand{\theenumi}{\Roman{enumi}}
\item
	The subgraph $\Hstart$ contains an initial node,
	whereas $\Hfake$ does not;
	both have one external edge in the direction $a$.
\item
	\label{lemma_H_start_H_fake__getting_out}
	There is an $n$-state automaton,
	which begins its computation on $\Hstart$ in the initial node,
	and leaves this subgraph by the external edge.
\item
	\label{lemma_H_start_H_fake__getting_in}
	Every automaton with fewer than $2(k-3)(n-1)$ states,
	having entered $\Hstart$ and $\Hfake$ by the external edge in the same state,
	either leaves both graphs in the same state,
	or accepts both,
	or rejects both,
	or loops on both.
\end{enumerate}
\end{lemma}

The proof reuses a graph constructed by the authors in a recent paper~\cite{MartynovaOkhotin_lb}.
Originally, it was used to show that there is an $n$-state graph-walking automaton,
such that every automaton that accepts the same graphs
and returns to the initial node after acceptance
must have at least $2(k-3)(n-1)$ states~\cite[Thm.~18]{MartynovaOkhotin_lb},
cf.\ upper bound $2nk+n$~\cite[Thm.~9]{MartynovaOkhotin_lb}.
A summary of the proof is included for completeness,
as well as adapted to match the statement of the lemma.

\begin{proof}[Summary of the proof.]
The graph is constructed in two stages.
First, there is a graph $G$ presented in Figure~\ref{f:returning_simplified},
with two long chains of nodes in the direction $\pm a$
connected by two bridges in the direction $\pm b$,
which are locally indistinguishable from loops by $\pm b$ at other nodes.
In order to get from the initial node $v_0$
to the node $\vexit$,
an $n$-state automaton counts up to $n-1$ to locate the left bridge,
then crosses the bridge and continues moving to the right.
The journey back from $\vexit$ to $v_0$
requires moving in the direction $-a$
in at least $n-1$ distinct states~\cite[Lemma~17]{MartynovaOkhotin_lb}.

\begin{figure}[t]
	\centerline{\includegraphics[scale=1]{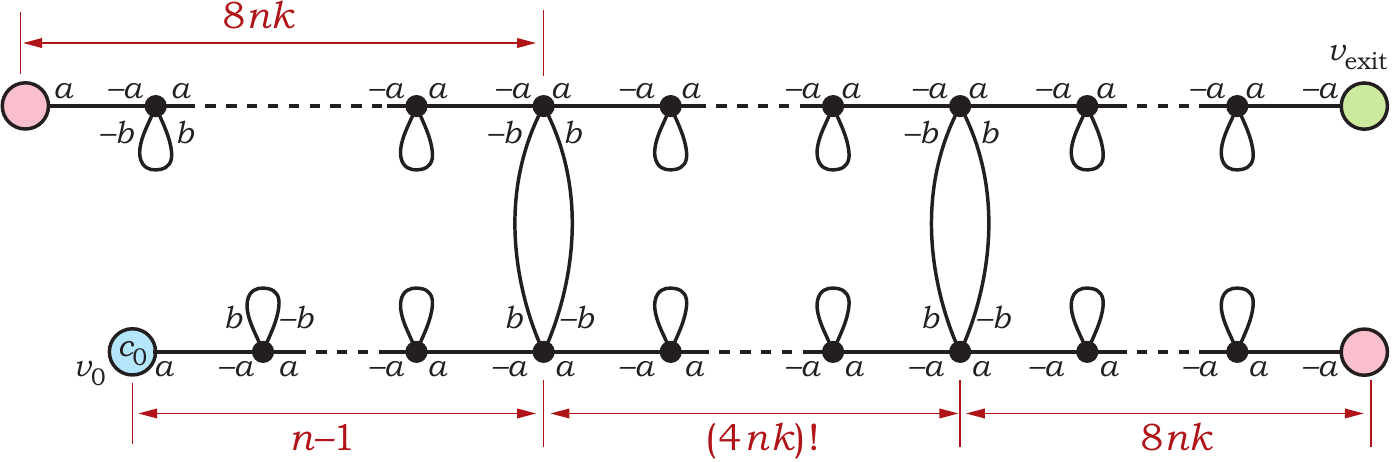}}
	\caption{The graph $G$.}
	\label{f:returning_simplified}
\end{figure}

In order to get a factor of $2(k-3)$,
another construction is used on top of this.
Every $(a, -a)$-edge in the horizontal chains
is replaced with a certain subgraph called a \emph{diode},
with $9(4nk)!+2$ nodes.
This subgraph is easy to traverse in the direction $a$:
an automaton can traverse it in a single state, guided by labels inside the diode,
so that the graph $G$ in Figure~\ref{f:returning_simplified},
with diodes substituted, can be traversed from $v_0$ to $\vexit$ using $n$ states.
However, as its name implies,
the diode is hard to traverse backwards:
for every state, in which the automaton finishes the traversal in the direction $-a$,
it must contain $2(k-3)-1$ extra states~\cite[Lemma~15]{MartynovaOkhotin_lb}.
Combined with the fact that there need to be at least $n-1$ states after moving by $-a$
for the automaton to get from $\vexit$ to $v_0$,
this shows that $2(k-3)(n-1)$ states are necessary to get from $\vexit$ to $v_0$
after the substitution of diodes.

Let $G_{diodes}$ be the graph in Figure~\ref{f:returning_simplified}, with diodes substituted.
It is defined over a signature with $k$ directions,
and among them the directions $\pm a$ and $\pm b$.
This signature is taken as $\Sstart$ in Lemma~\ref{lemma_H_start_H_fake}.

The graph $\Hstart$ is defined by removing the node $\vexit$ from $G_{diodes}$,
and the edge it was connected by
becomes an external edge in the direction $a$.
The other graph $\Hfake$ is obtained by relabelling the initial node $v_0$,
so that it is no longer initial.
An $n$-state automaton that gets out of $\Hstart$ has been described above.

Every automaton that enters $\Hstart$ or $\Hfake$ from the outside
needs at least $2(k-3)(n-1)$ states to get to $v_0$,
because returning from $\vexit$ to $v_0$ on $G_{diodes}$
requires this many states.
Then, an automaton with fewer states never reaches $v_0$,
and thus never encounters any difference between these subgraphs.
Thus, it carries out the same computation on both subgraphs $\Hstart$ and $\Hfake$,
with the same result.
\end{proof}

Now, using the subgraphs $\Hstart$ and $\Hfake$ as building blocks,
the next goal is to construct a subgraph which encodes a number from $0$ to $n-1$,
so that this number is easy to calculate along with getting out of this subgraph for the first time,
but if it is ever forgotten, then it cannot be recovered without using too many states.
For each number $i \in \{0,\ldots, n-1\}$ and for each direction $d \in D$,
this is a graph $F_{i,d}$
that contains the initial label and encodes the number $i$,
and a graph $F_d$ with no initial label that encodes no number at all.

\begin{lemma}\label{lemma_F_i_d}
For every $k \geqslant 4$ there is a signature $S_F$
obtained from $\Sstart$
by adding several new node labels,
such that, for every $n \geqslant 2$
there are subgraphs $F_{i,d}$ and $F_d$,
for all $i \in \{0, \ldots, n-1\}$ and $d \in D$,
with the following properties.
\begin{enumerate}\renewcommand{\theenumi}{\Roman{enumi}}
\item
Each subgraph $F_{i,d}$ and $F_d$ has one external edge in the direction $d$.
Subgraphs of the form $F_{i,d}$ have an initial node,
and subgraphs $F_d$ do not have one.

\item
	\label{lemma_F_i_d__getting_out}
	There is an automaton with states $\{q_0, \ldots, q_{n-1}\}$,
	which, having started on every subgraph $F_{i,d}$ in the initial node,
	eventually gets out in the state $q_i$.

\item
	\label{lemma_F_i_d__getting_in}
	Every automaton with fewer than $2(k-3)(n-1)$ states,
	having entered $F_{i,d}$ and $F_d$ with the same $d$
	by the external edge in the same state,
	either leaves both subgraphs in the same state,
	or accepts both,
	or rejects both,
	or loops on both.
\end{enumerate}
\end{lemma}

\begin{figure}[t]
	\centerline{\includegraphics[scale=1.1]{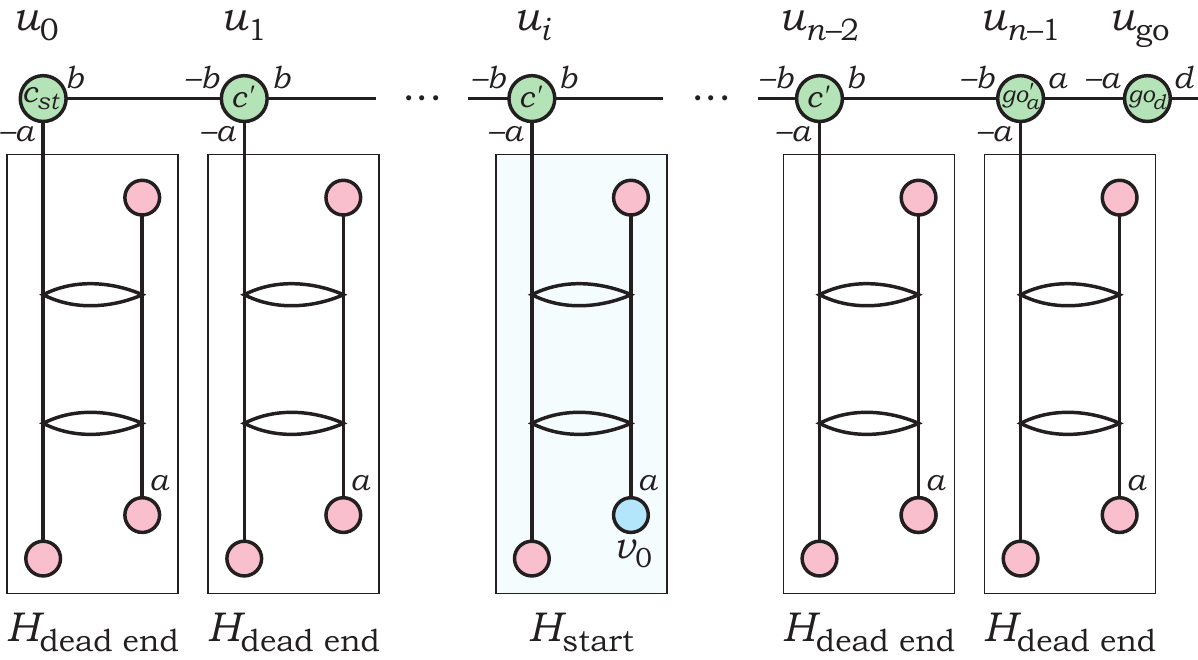}}
	\caption{The subgraph $F_{i,d}$, with $d \neq -a$;
		for $d=-a$ the subgraph has $u_{n-1}$ labelled with $go'_b$,
		and a $(b, -b)$-edge to $u_{go}$.}
	\label{f:F_i_d}
\end{figure}

Each subgraph $F_{i,d}$ is a chain of $n$ nodes,
with the subgraph $\Hstart$ attached at the $i$-th position,
and with $n-1$ copies of $\Hfake$ attached at the remaining positions,
as illustrated in Figure~\ref{f:F_i_d}.
The automaton in Part~\ref{lemma_F_i_d__getting_out} gets out of $\Hstart$
and then moves along the chain to the left,
counting the number of steps,
so that it gets out of the final node $u_{go}$ in the state $q_i$.
The proof of Part~\ref{lemma_F_i_d__getting_in}
relies on Lemma~\ref{lemma_H_start_H_fake} (part~\ref{lemma_H_start_H_fake__getting_in}):
if an automaton enters $F_{i,d}$ and $F_d$ from the outside,
it ends up walking over the chain
and every time it enters any of the attached subgraphs $\Hstart$ and $\Hfake$,
it cannot distinguish between them and continues in the same way on all $F_{i,d}$ and $F_d$.

\begin{proof}
The new signature $S_F$ has the following new non-initial node labels:
$\{c_{st}, c', go'_a, go'_b\}\cup \set{go_d}{d \in D}$. 
The labels have the following sets of directions:
$D_{c_{st}} = \{-a, b\}$, $D_{c'} = \{-a,-b,b\}$, $D_{go'_a} = \{-a,-b,a\}$, 
$D_{go'_b} = \{-a,-b,b\}$, $D_{go_d} = \{-a,d\}$ with $d \neq -a$, 
and $D_{go_{-a}} = \{-b,-a\}$.

For $n \geqslant 2$, the subgraphs $F_{i,d}$ and $F_d$
are constructed as follows,
using the subgraphs $\Hstart$ and $\Hfake$ given in Lemma~\ref{lemma_H_start_H_fake}.

The subgraph $F_{i,d}$, illustrated in Figure~\ref{f:F_i_d},
is a chain of nodes $u_0$, \ldots, $u_{n-1}, u_{go}$;
the first $n-1$ nodes are linked with $(b, -b)$-edges.
The node $u_{n-1}$ is linked to $u_{go}$ by an edge $(a, -a)$ if $d \neq -a$,
and by an edge $(b, -b)$ for $d=-a$.
The label of $u_0$ is $c_{st}$, 
the nodes $u_1, \ldots, u_{n-2}$ all have label $c'$, 
and $u_{n-1}$ is labelled with $go'_a$, if $d \neq -a$, 
or with $go'_b$, if $d = -a$.
The node $u_{go}$ has label $go_d$,
and has an external edge in the direction $d$.

Each node $u_0, \ldots, u_{n-1}$ has a subgraph $\Hstart$ or $\Hfake$
attached in the direction $-a$.
This is $\Hstart$ for $u_i$, and $\Hfake$ for the rest of these nodes.

The subgraph $F_d$ is the same as $F_{i,d}$, 
except for having $\Hfake$ attached to all nodes $u_0, \ldots, u_{n-1}$.

It is left to prove that the subgraphs $F_{i,d}$ and $F_d$ thus constructed
satisfy the conditions in the lemma.

Part~\ref{lemma_F_i_d__getting_out} of this lemma
asserts that there is an $n$-state automaton
that gets out of $F_{i,d}$ in the state $q_i$,
for all $i$ and $d$.
Having started in the initial node inside a subgraph $\Hstart$,
the automaton operates as the $n$-state automaton
given in Lemma~\ref{lemma_H_start_H_fake}(part \ref{lemma_H_start_H_fake__getting_out})
which leaves $\Hstart$ in some state $q$.
Denote this state by $q_{n-1}$, and let $\{q_0, \ldots, q_{n-2}\}$ be the remaining states
(it does not matter which of these states is initial).
Then the automaton follows the chain of nodes to the right,
decrementing the number of the state at each node labelled with $c_{st}$ or $c'$.
At the nodes labelled with $go'_a$, $go'_b$ or $go_d$,
the automaton continues to the right without changing its state.
Thus, for each subgraph $F_{i,d}$, the automaton gets out in the state $q_i$, as desired.

Turning to the proof of Part~\ref{lemma_F_i_d__getting_in},
consider an automaton with fewer than $2(k-3)(n-1)$ states
and let $d \in D$ be any direction.
The subgraphs $F_{i,d}$ for various $i$, as well as the subgraph $F_d$,
differ only in the placement of the subgraph $\Hstart$
among the subgraphs $\Hfake$, or in its absense.
On each of the subgraphs $F_{i,d}$ or $F_d$,
the automaton first moves over the chain of nodes $u_0, \ldots, u_{n-1}, u_{go}$,
which is the same in all subgraphs.
Whenever, at some node $u_j$, it enters the $j$-th attached subgraph,
whether it is $\Hstart$ or $\Hfake$,
according to Lemma~\ref{lemma_H_start_H_fake},
it is not able to distinguish between them,
and the computation has the same outcome:
it either emerges out of each of the attached subgraphs in the same state,
or accepts on either of them, etc.
If the computation continues,
it continues from the same state and the same node in all $F_{i,d}$ and $F_d$,
and thus the computations on all these subgraphs proceed in the same way
and share the same outcome.
\end{proof}

\begin{proof}[Proof of Theorem~\ref{theorem_inverse_homomorphism_lower_bound}]
The signature $S$ is defined by adding some further node labels
to the signature $S_F$ from Lemma~\ref{lemma_F_i_d},
maintaining the same set of directions $D$.
Let the directions be cyclically ordered,
with $next(d)$ representing the next direction after $d$
according to this order,
whereas $prev(d)$ is the previous direction.
The order is chosen so that, for each direction $d$,
its opposite direction is neither $next(d)$ nor $next(next(d))$.

The new node labels, all non-initial,
are:
$\set{go_{-d,a}}{d \in D \setminus \{-a\}} \cup \{go_{a,b}, c_{-}, q_0?\} 
\cup \set{d?}{d \in D} \cup \set{acc_d, rej_d}{d \in D}$. 
These labels have the following sets of directions:
$D_{go_{d_1, d_2}} = \{d_1,d_2\}$; 
$D_{c_{-}} = \{-a,a\}$; $D_{q_0?} = \{-a\}$; $D_{d?} = D$ for all $d \in D$; 
$D_{acc_d} = D_{rej_d} = \{-d, -next(d), next(next(d))\}$ for all $d \in D$,
where the directions $-d, -next(d), next(next(d))$ are pairwise distinct by assumption.

The node-replacement homomorphism $h$ mapping graphs over $S$ to graphs over $S$
affects only labels of the form $d?$, with $d \in D$,
whereas the rest of the labels are unaffected,
that is, mapped to single-node subgraphs with the same label.
Each label $d?$, for $d \in D$,
is replaced with a circular subgraph $h(d?)$
as illustrated in Figure~\ref{f:G_i_d_dprime}.
Its nodes are $v_e$, for all $e \in D$.
The node $v_d$ has label $acc_d$,
and every node $v_e$, with $e \neq d$, is labelled with $rej_e$. 
Each node $v_e$, with $e \in D$,
is connected to the next node $v_{next(e)}$ by an edge in the direction $next(next(e))$;
also it has an external edge in the direction $-e$.
Overall, the subgraph $h(d?)$ has an external edge in each direction,
as it should have, since $D_{d?} = D$. 

\begin{figure}[t]
	\centerline{\includegraphics[scale=1.1]{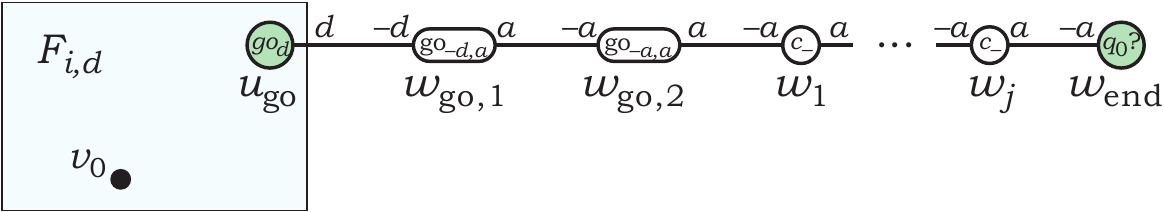}}
	\caption{The graph $G_{i,j,d}$, with $d \neq -a$;
		for $d=-a$ the graph has $w_{go,1}$ labelled with $go_{a,b}$
		and $w_{go,2}$ labelled with $go_{-b,a}$,
		linked with a $(b, -b)$-edge.}
	\label{f:G_i_j_d}
\end{figure}

The graph $G_{i,j,d}$
is defined by taking $F_{i,d}$ from Lemma~\ref{lemma_F_i_d}
and attaching to it a chain of $j+3$ nodes,
as shown in Figure~\ref{f:G_i_j_d}.
The new nodes are denoted by $w_{go,1}, w_{go,2}, w_1, \ldots, w_j, w_{end}$,
where the external edge of $F_{i,d}$ is linked to $w_{go,1}$ in the direction $d$.
If $d \neq -a$,
then the nodes $w_{go,1}$ and $w_{go,2}$
have labels $go_{-d,a}$ and $go_{-a,a}$,
and are connected with an $(a,-a)$-edge;
and if $d=-a$, then the labels are $go_{a,b}$ and $go_{-b,a}$,
and the edge is $(b, -b)$.
The nodes $w_1, \ldots, w_j$ are labelled with $c_{-}$,
the label of $w_{end}$ is $q_0?$,
and all of them are connected with $(a, -a)$-edges.

The form of the graph $G_{i,d,d'}$,
presented in Figure~\ref{f:G_i_d_dprime} for the case $d=d'$, is simpler.
It has a subgraph $F_{i,d}$ with the initial node,
and $k-1$ subgraphs $F_e$, with $e \in D \setminus \{d\}$.
The external edges of these $k$ subgraphs
are all linked to a new node $v$ labelled with $d'?$.

\begin{figure}[t]
	\centerline{\includegraphics[scale=1.1]{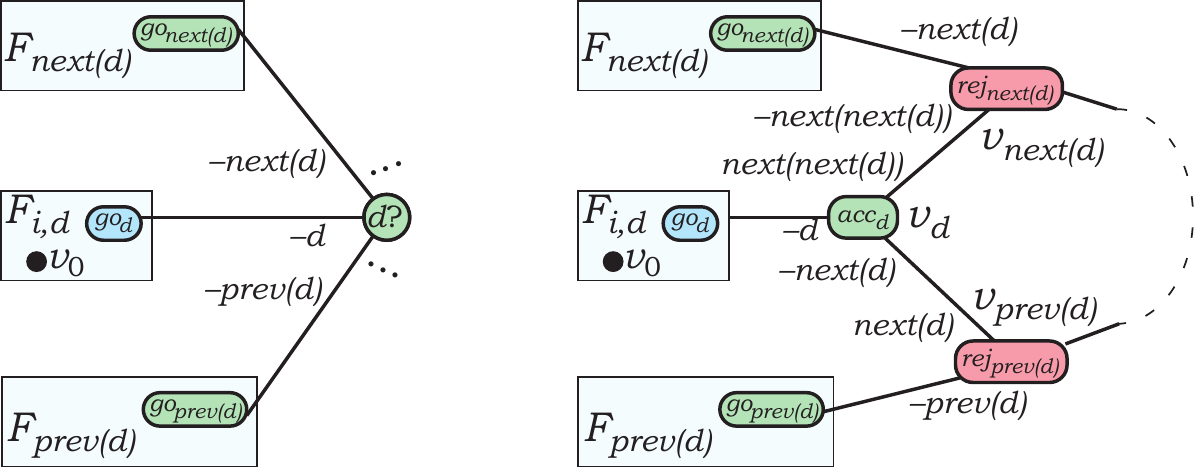}}
	\caption{The graph $G_{i,d,d}$ and its image $h(G_{i,d,d})$.}
	\label{f:G_i_d_dprime}
\end{figure}

\begin{ourclaim}\label{theorem_inverse_homomorphism_lower_bound__A_claim}
There exists an $n$-state automaton $A$,
which accepts $h(G_{i,j,d})$ if and only if $i=j$,
and which accepts $h(G_{i,d,d'})$ if and only if $d=d'$.
\end{ourclaim}
\begin{proof}
The automaton is based on the one
defined in Lemma~\ref{lemma_F_i_d} (part~\ref{lemma_F_i_d__getting_out}).
It works over the signature $S_F$
and has $n$ states $\{q_0, \ldots, q_{n-1}\}$.
Having started on a graph $F_{i,d}$, it eventually gets out in the state $q_i$. 
It remains to define the right transitions
by the new labels in the signature $S$.
At each label $go_{d_1,d_2}$, the automaton moves in the direction $d_2$ in the same state.
At a label $c_{-}$ the automaton decrements the number of its current state and moves in the direction $a$.
If it ever comes to a label $c_-$ in the state $q_0$, it rejects.
At the label $q_0?$, the automaton accepts if its current state is $q_0$,
and rejects in all other states.
Turning to the labels introduced by the homomorphism,
for all $d \in D$,
the automaton immediately accepts at $acc_d$ and rejects at $rej_d$,
regardless of its current state.

To see that the automaton $A$ operates as claimed,
first consider its computation on the graph $h(G_{i,j,d}) = G_{i,j,d}$.
It starts at the initial node in $F_{i,d}$,
then leaves $F_{i,d}$ in the state $q_i$,
passes through the nodes $w_{go,1}$ and $w_{go,2}$ without changing its state,
and then decrements the number of the state
at the nodes $w_1, \ldots, w_j$.
If $i = j$, then the automaton $A$ makes $j$ decrementations,
and arrives to the node with the label $q_0?$ in the state $q_0$, and accordingly accepts.
If $i>j$, then it comes to $q_0?$ in the state $q_{i-j} \neq q_0$ and rejects.
If $i<j$, then $A$ enters the state $q_0$ at one of the labels $c_{-}$, 
and rejects there.
Thus, $A$ works correctly on graphs of the form $h(G_{i,j,d})$.

In the graph $h(G_{i,d,d'})$, the homomorphism has replaced the node $v$ from $G_{i,d,d'}$
with a ring of nodes with labels $acc_{d'}$ and $rej_e$, with $e \neq d'$.
The automaton $A$ starts in the subgraph $F_{i,d}$
and leaves it in the direction $d$,
thus entering the ring at the node $v_d$.
Then, if $d = d'$, it sees the label $acc_{d'}$ and accepts,
and otherwise it sees the label $rej_d$ and rejects.
The automaton does not move along the circle.
\end{proof}

The automaton is based on the one
defined in Lemma~\ref{lemma_F_i_d} (part~\ref{lemma_F_i_d__getting_out}).
On the graph $h(G_{i,j,d})$,
it gets out of the subgraph $F_{i,d}$ in the state $q_i$,
and then decrements the counter $j$ times as it continues to the right;
if it reaches the end of the chain in $q_0$, it accepts.
On the graph $h(G_{i,d,d'})$, the automaton comes to the ring $h(d'?)$;
if $d=d'$, it arrives at the node with label $acc_d$ and accepts;
otherwise, the label is $rej_d$, and it rejects.

\begin{ourclaim}
Let an automaton $B$ accept a graph $G$
if and only if $A$ accepts $h(G)$.
Then $B$ has at least $nk$ states.
\end{ourclaim}

The proof is by contradiction.
Suppose that $B$ has fewer than $nk$ states.
Since
$nk \leqslant 2\cdot\frac{2}{3}k\cdot\frac{3}{4}n \leqslant 2(k-3)(n-1)$,
Lemma~\ref{lemma_F_i_d} (part~\ref{lemma_F_i_d__getting_in}) applies,
and the automaton $B$ cannot distinguish between the subgraphs $F_{i,d}$ and $F_d$
if it enters them from the outside.

On the graph $G_{i,j,d}$,
the automaton must check that $i$ is equal to $j$,
where the latter is the number of labels $c_{-}$ after the exit from $F_{i,d}$.
In order to check this, $B$ must exit this subgraph.
Denote by $q_{i,d}$
the state, in which the automaton $B$ leaves the subgraph $F_{i,d}$ for the first time.
There are $nk$ such states $\set{q_{i,d}}{i = 0, \ldots, n-1; d \in D}$,
and since $B$ has fewer
states, some of these states must coincide.
Let $q_{i,d} = q_{j,d'}$, where $d \neq d'$ or $i \neq j$.
There are two cases to consider.

\begin{itemize}
\item
Case 1: $d \neq d'$.
The automaton $B$ must accept $G_{i,d,d}$ and reject $G_{j,d',d}$. 
On either graph, it first arrives to the corresponding node $v$
in the same state $q_{i,d} = q_{j,d'}$,
without remembering the last direction taken.
Then, in order to tell these graphs apart,
the automaton must carry out some further checks.
However, every time $B$ leaves the node $v$ in any direction $e \in D$,
it enters a subgraph, which is either the same 
in $G_{i,d,d}$ and $G_{j,d',d}$ (if $e \neq d,d'$),
or it is a subgraph that is different in the two graphs,
but, according to Lemma~\ref{lemma_F_i_d} (part~\ref{lemma_F_i_d__getting_in}),
no automaton of this size can distinguish between these subgraphs.
Therefore, $B$ either accepts both graphs,
or rejects both graphs,
or loops on both, which is a contradiction.

\item
Case 2: $d = d'$ and $i \neq j$.
In this case, consider the computations of $B$ 
on the graphs $G_{i,j,d}$ and $G_{j,j,d}$:
the former must be rejected, the latter accepted.
However, by the assumption, the automaton leaves $F_{i,d}$ and $F_{j,d}$ 
in the same state $q_{i,d} = q_{j,d}$.
From this point on, the states of $B$ in the two computations
are the same while it walks outside of $F_{i,d}$ and $F_{j,d}$,
and each time it reenters these subgraphs,
by Lemma~\ref{lemma_F_i_d} (part~\ref{lemma_F_i_d__getting_in}),
it either accepts both, or rejects both, or loops on both,
or leaves both in the same state.
Thus, the whole computations have the same outcome,
which is a contradiction.
\end{itemize}

The contradiction obtained shows that $B$ has at least $nk$ states.
\end{proof}

\section{A characterization of regular tree languages}\label{section_reg_tree}

The next question investigated in this paper
is whether the family of graph languages recognized by graph-walking automata
is closed under homomorphisms.
In this section, non-closure is established already for tree-walking automata
and for injective homomorphisms.

The proof is based on a seemingly unrelated result.
Consider the following known representation of regular string languages.

\begin{oldtheorem}[Latteux and Leguy~\cite{LatteuxLeguy}]\label{reg_as_hminus1_h_hminus1_c_theorem}
For every regular language $L \subseteq \Sigma^*$
there exist alphabets $\Omega$ and $\Gamma$,
a special symbol $\#$,
and homomorhisms $f \colon \Omega^* \to \#^*$,
$g \colon \Omega^* \to \Gamma^*$
and $h \colon \Sigma^* \to \Gamma^*$,
such that $L=h^{-1}(g(f^{-1}(\#))$.
\end{oldtheorem}

A similar representation shall now be established for regular tree languages,
that is, those recognized by deterministic bottom-up tree automata.

For uniformity of notation, tree and tree-walking automata
shall be represented in the notation of graph-walking automata, 
as in Section~\ref{section_definitions},
which is somewhat different from the notation used in the tree automata literature.
This is only notation, and the trees and the automata are mathematically the same.

\begin{definition}
A signature $S = (D, -, \Sigma, \Sigma_0, (D_a)_{a \in \Sigma})$
is a \emph{tree signature}, if it is of the following form.
The set of directions is $D=\{+1, -1, \ldots, +k, -k\}$, for some $k \geqslant 1$,
where directions $+i$ and $-i$ are opposite to each other. 
For every label $a \in \Sigma$,
the number of its children is denoted by $\rank a$, with $0 \leqslant \rank a \leqslant k$.
Every initial label $a_0 \in \Sigma_0$ has directions $D_{a_0}=\{+1, \ldots, +\rank a_0\}$.
Every non-initial label $a \in \Sigma \setminus \Sigma_0$ 
has the set of directions $D_a=\{-d, +1, \ldots, +\rank a\}$,
for some $d \in \{1, \ldots, k\}$.

A tree is a connected graph over a tree signature.
\end{definition}

This definition implements the classical notion of a tree as follows.
The initial node is the root of a tree.
In a node $v$ with label $a$, the directions $\{+1, \ldots, +\rank a\}$ lead to its children.
The child in the direction $+i$ accordingly has direction $-i$ to its parent.
This direction to the parent is absent in the root node.
Labels $a$ with $\rank a = 0$ are used in the leaves.

\begin{definition}
A (deterministic bottom-up) tree automaton
over a tree signature $S = (D, -, \Sigma, \Sigma_0, (D_a)_{a \in \Sigma})$
is a triple $A = (Q, q_{acc}, (\delta_a)_{a\in \Sigma})$, where
\begin{itemize}
\item
	$Q$ is a finite set of states;
\item
	$q_{acc} \in Q$ is the accepting state, effective in the root node;
\item
	$\delta_a \colon Q^{\rank a} \to Q$, for each $a \in \Sigma$,
	is a function computed at the label $a$.
	If $\rank a  = 0$, then $\delta_a$ is a constant that sets the state in a leaf.
\end{itemize}

Given a tree $T$ over a signature $S$,
a tree automaton $A$
computes the state in each node, bottom-up.
The state in each leaf $v$ labelled with $a$
is set to be the constant $\delta_a()$.
Once a node $v$ labelled with $a$
has the states in all its children computed
as $q_1, \ldots, q_{\rank a}$, 
the state in the node $v$ is computed as $\delta_a(q_1, \ldots, q_{\rank a})$.
This continues until the value in the root is computed.
If it is $q_{acc}$, then the tree is accepted, and otherwise it is rejected.
The tree language recognized by $A$
is the set of all trees over $S$ that $A$ accepts.
A tree language is called regular if it is recognized by some tree automaton.
\end{definition}

The generalization of Theorem~\ref{reg_as_hminus1_h_hminus1_c_theorem}
to the case of trees
actually uses only two homomorphisms, not three.
The inverse homomorphism $f^{-1}$ in Theorem~\ref{reg_as_hminus1_h_hminus1_c_theorem}
is used to generate the set of all strings with a marked first symbol
out of a single symbol.
Trees cannot be generated this way.
The characterization given below starts from the set of all trees over a certain signature,
in which the root is already marked by definition;
this achieves the same effect as $f^{-1}\big(\{\#\}\big)$ in Theorem~\ref{reg_as_hminus1_h_hminus1_c_theorem}.
The remaining two homomorphisms do basically the same as in the original result,
only generalized to trees.

\begin{theorem}\label{theorem_tree_languages_homomorphic_form}
Let $L$ be a regular tree language over some tree signature $S_{reg}$.
Then there exist tree signatures $S_{comp}$ and $S_{mid}$,
and injective homomorphisms $g \colon L(S_{comp}) \to L(S_{mid})$ 
and $h \colon L(S_{reg}) \to L(S_{mid})$,
such that $L = h^{-1}(g(L(S_{comp})))$.
\end{theorem}
\begin{proof}
The signature $S_{mid}$ extends $S_{reg}$
with a few new non-initial node labels;
the set of directions is preserved.
The new labels are $k$ labels for internal nodes, $e_1, \ldots, e_k$,
with $\rank e_i = k$ and $D_{e_i} = \{-i, +1, \ldots, +k\}$,
and $k$ more labels for leaves,
$end_1, \ldots, end_k$,
with $\rank end_i=0$ and $D_{end_i}=\{-i\}$.
These labels are used to construct a \emph{fishbone subgraph}:
a fishbone subgraph of length $\ell$ in the direction $i$
is a chain of $\ell$ internal nodes, all labelled with $e_i$,
which begins and ends with external edges in the directions $-i$ and $+i$;
all directions except $\pm i$ lead to leaves labelled with $end_j$.

\begin{figure}[t]
	\centerline{\includegraphics[scale=1.1]{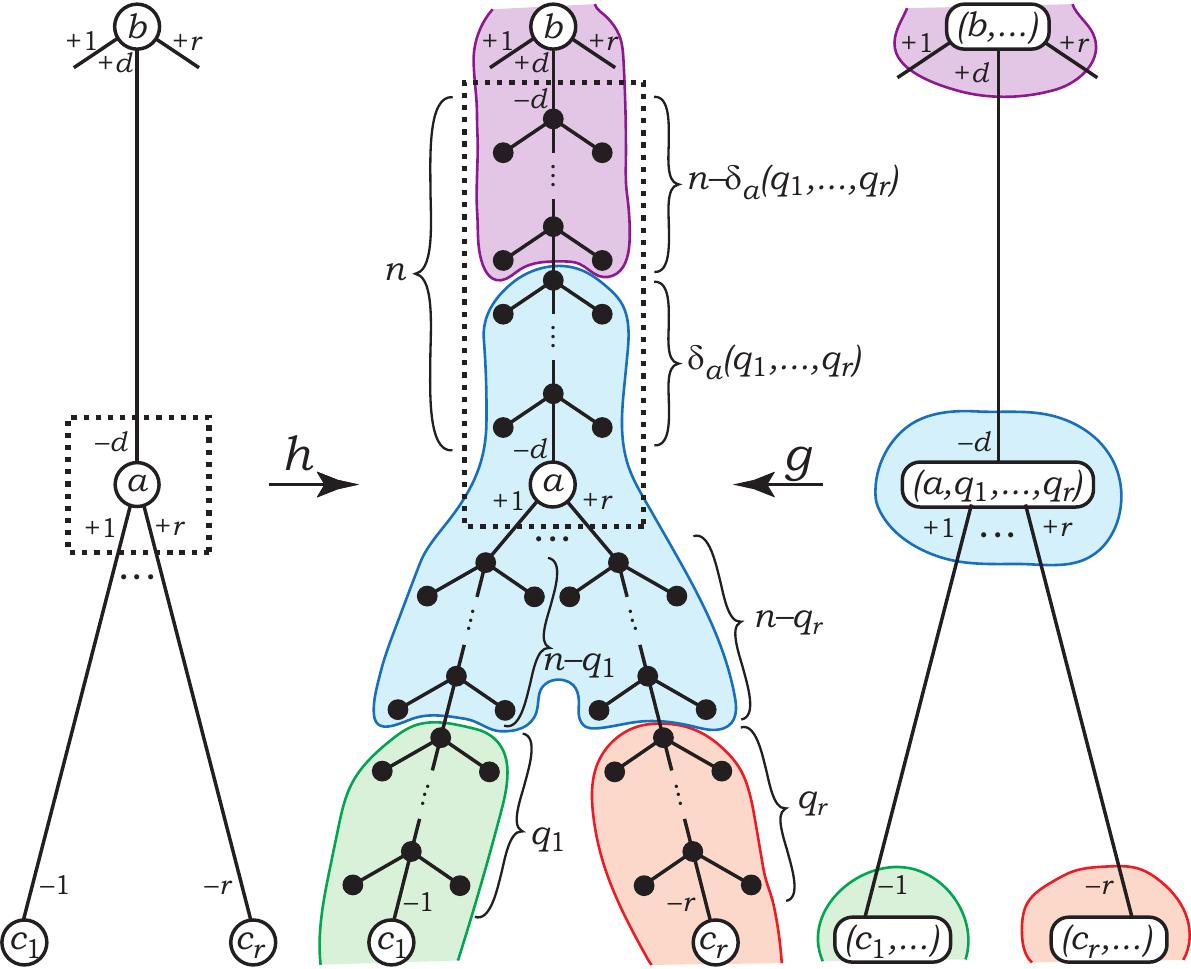}}
	\caption{Homomorphisms $h$ and $g$ mapping the original tree $T$ (left)
		and the corresponding valid annotated tree $T_{comp}$ (right)
		to the same tree with fishbones.}
	\label{f:regular_tree_h_and_g}
\end{figure}

An injective homomorphism $h \colon L(S_{reg}) \to L(S_{mid})$
is defined to effectively replace each $(+i, -i)$-edge
with a fishbone subgraph of length $n$ in the direction $i$,
without affecting the original nodes and their labels,
as illustrated in Figure~\ref{f:regular_tree_h_and_g}.
Formally, $h$ replaces each non-initial node
labelled with $a \in \Sigma \setminus \Sigma_0$
as follows.
Let $D_a = \{-d,+1, \ldots, +\rank a\}$ be its set of directions.
Then, $h(a)$ is the following subgraph:
it consists of a node with the same label $a$,
a fishbone subgraph of length $n$ attached in the direction $-d$,
and $\rank a$ external edges in the directions $+1, \ldots, +\rank a$.
The initial node is mapped to itself.

The main idea of the construction
is to take a tree accepted by $A$ and annotate node labels
with the states in the accepting computation of $A$ on this tree.
Another homomorphism $g$ maps such annotated trees
to trees over the signature $S_{mid}$, with fishbones therein.
Annotated trees that correctly encode a valid computation
are mapped to trees with all fishbones of length exactly $n$;
then, $h^{-1}$ decodes the original tree out of this encoding.
On the other hand, any mistakes in the annotation
are mapped by $g$ to a tree with some fishbones of length other than $n$,
and these trees have no pre-images under $h$.

Trees with annotated computations are defined over the signature $S_{comp}$.
This signature uses the same set of directions as in $S_{reg}$.
For every non-initial label $a \in \Sigma \setminus \Sigma_0$ in $S_{reg}$,
the signature $S_{comp}$ has $|Q|^{\rank a}$ different labels
corresponding to all possible vectors of states in its children.
Thus, for every $\bm{q} = (q_1, \ldots, q_{\rank a}) \in Q^{\rank a}$,
there is a non-initial label $(a,\bm{q})$ in $S_{comp}$,
with $\rank (a, \bm{q}) = \rank a$ and $D_{(a,\bm{q})} = D_a$.
For every initial label $a_0 \in \Sigma_0$ in $S_{reg}$,
the signature $S_{comp}$ contains only those initial labels $(a_0,\bm{q})$, 
for which the vector $\bm{q} \in Q^{\rank a_0}$ of states in the children leads to acceptance,
that is, $\delta_{a_0}(\bm{q}) = q_{acc}$.
The rank and the set of directions are also inherited:
$\rank (a_0,\bm{q}) = \rank a_0$ and $D_{(a_0,\bm{q})} = D_{a_0}$.
There is at least one initial label in $S_{comp}$, because $L \neq \emptyset$.
If $\rank a = 0$, then the set $Q^{\rank a}$ 
contains a unique vector $\bm{q}$ of length $0$.
Such a label has only one copy $(a,\bm{q})$ in the signature $S_{comp}$,
or none at all, if $a = a_0 \in \Sigma_0$ and $\delta_a(\bm{q}) \neq q_{acc}$.

For every tree $T$ over $S_{reg}$ that is accepted by $A$,
the accepting computation of $A$ on $T$
is represented by a tree $T_{comp}$ over the signature $S_{comp}$,
in which every label is annotated with the vector of states in the children of this node.
Annotated trees that do not encode a valid computation
have a mismatch in at least one node $v$,
that is, the state in some $i$-th component of the vector in the label
does not match the state computed in the $i$-th child.
It remains to separate valid annotated trees from invalid ones.

The homomorphism $g \colon L(S_{comp}) \to L(S_{mid})$ is formally defined as follows.
Let $(a,\bm{q})$ be a non-root label with $\bm{q} = (q_1, \ldots, q_{\rank a}) \in Q^{\rank a}$
and $D_a = \{-d, +1, \ldots, +\rank a\}$.
Then, $g$ maps $(a,\bm{q})$ to a subgraph $g((a,\bm{q}))$,
which is comprised of a central node $v_{center}$ labelled with $a$,
with fishbone subgraphs attached in all directions.
The direction $-d$ is attached to the bottom of a fishbone graph
in the direction $d$ of length $\delta_a(\bm{q})$.
The subgraph attached in each direction $+i$
is a fishbone of length $n-q_i$ in the direction $i$.
The external edges of the subgraph $g((a,\bm{q}))$ come out of these fishbones.
If $n-q_i=0$, then the fishbone of length $0$
is an external edge in the direction $+i$.
The image of a root label $(a_0, \bm{q})$ under $g$ is defined in the same way,
except for not having a direction $-d$ and the corresponding fishbone.

Images of trees under the homomorhism $g$ are of the following form.

\begin{ourclaim}\label{theorem_tree_languages_homomorphic_form__g_claim}
Let $\widetilde{T}$ be an annotated tree over the signature $S_{comp}$,
with the nodes $v^1, \ldots, v^m$
labelled with $(a^1, \bm{q^1}), \ldots, (a^m, \bm{q^m})$.
Then the tree $g(\widetilde{T})$ is obtained from $\widetilde{T}$ as follows:
every label $(a^t, \bm{q^t})$ is replaced with $a^t$,
and every edge $(+i,-i)$ linking a parent $v^s$ to a child $v^t$ in $\widetilde{T}$
is replaced with a fishbone of length $n-q^s_i+\delta_{a^t}(\bm{q^t})$ in the direction $i$.
\end{ourclaim}

The image of all valid annotated trees under $g$ is exactly $h(L)$.

\begin{ourclaim}
Let $T$ be a tree accepted by $A$,
and let $T_{comp}$ be an annotated tree that encodes the computation of $A$ on $T$.
Then, the homomorphism $g$ maps $T_{comp}$ to $h(T)$.
\end{ourclaim}

Indeed, if an annotated tree represents a valid computation,
then, in Claim~\ref{theorem_tree_languages_homomorphic_form__g_claim},
$q^s_i=\delta_{a^t}(\bm{q^t})$ holds
for every pair of a parent $v_s$ and its $i$-th child $v_t$,
and thus all fishbones are of length $n$, as in $h(T)$.
For the same reason, $g$ maps invalid annotated trees to trees without pre-images under $h$.
Therefore, $h^{-1}(g(L(S_{comp}))) = L$.

The homomorphism $h$ is injective, because it does not affect the node labels
and only attaches fixed subgraphs to them.
On the other hand, $g$ erases the second components of labels,
and its injectivity requires an argument.

\begin{ourclaim}\label{claim_g_injection}
The homomorphism $g$ is injective.
\end{ourclaim}
\begin{proof}
Let $T$ and $T'$ be trees over $S_{comp}$ that are mapped to the same tree $g(T)=g(T')$.
It is claimed that $T=T'$.
By Claim~\ref{theorem_tree_languages_homomorphic_form__g_claim},
both trees $T$ and $T'$ have the same set of nodes
and the same edges between these nodes,
as well as the same first components of their labels.

It remains to show that the second components of labels
at the corresponding nodes of $T$ and $T'$ also coincide.
This is proved by induction, from leaves up to the root.
For a leaf, the second component is an empty vector in both trees.
For every internal node $v^s$ in these trees,
let $(a^s, \bm{q^s})$ be its label in $T$
and let $(a^s, \bm{r^s})$ be its label in $T'$.
Consider its $i$-th child $v^t$;
by the induction hypothesis it has the same label $(a^t, \bm{q^t})$ in both trees.
Claim~\ref{theorem_tree_languages_homomorphic_form__g_claim}
asserts that the fishbone between $v_s$ and $v_t$ in $g(T)$
is of length $n-q^s_i+\delta_{a^t}(\bm{q^t})$,
and the length of the fishbone between $v_s$ and $v_t$ in $g(T')$
is $n-r^s_i+\delta_{a^t}(\bm{q^t})$.
Since this is actually the same fishbone,
this implies that $q^s_i=r^s_i$,
and the labels of $v_s$ in both trees are equal.
This completes the induction step and proves that $T=T'$.
\end{proof}

Thus, the homomorphisms $h$ and $g$ are as desired.
\end{proof}

\begin{theorem}
The class of tree languages recognized by tree-walking automata
is not closed under injective homomorphisms.
\end{theorem}
\begin{proof}
Suppose it is closed.
It is claimed that then every regular tree language
is recognized by a tree-walking automaton.
Let $L$ be a regular tree language
over some tree signature $S_{reg}$.
Then, by Theorem~\ref{theorem_tree_languages_homomorphic_form},
there exist tree signatures $S_{comp}$ and $S_{mid}$,
and injective homomorphisms $g \colon L(S_{comp}) \to L(S_{mid})$ 
and $h \colon L(S_{reg}) \to L(S_{mid})$,
such that $L = h^{-1}(g(L(S_{comp})))$.
The language $L(S_{comp})$ is trivially recognized by a tree-walking automaton
that accepts every tree right away.
Then, by the assumption on the closure under $g$,
the language $g(L(S_{comp}))$ is recognized by another tree-walking automaton.
By Theorem~\ref{theorem_inverse_homomorphism_upper_bound},
its inverse homomorphic image $L$ is recognized by a tree-walking automaton as well.
This contradicts the result by Boja\'nczyk and Colcombet~\cite{BojanczykColcombet_reg}
on the existence of regular tree languages
not recognized by any tree-walking automata.
\end{proof}

\section{Future work}

The lower bound on the complexity of inverse homomorphisms
is obtained using graphs with cycles.
So it does not apply to the important case of tree-walking automata (TWA).
On the other hand, in the even more restricted case of two-way finite automata (2DFA),
the state complexity of inverse homomorphisms is known to be $2n$~\cite{JiraskovaOkhotin_2dfa},
which is in line of the $kn$ bound in this paper, as 2DFA have $k=2$.
It would be interesting to fill in the missing case of TWA.

Also, other recent lower bounds on the size of graph-walking automata~\cite{MartynovaOkhotin_lb}
do not apply to TWA, and require a separate investigation.

\end{document}